\documentclass[conference]{IEEEtran}
\ifCLASSINFOpdf
\else
\fi
\usepackage{amsmath,amssymb,amsfonts}
\usepackage{subfig}
\usepackage{cite}
\usepackage{algorithm}
\usepackage{graphicx}
\usepackage{enumerate}
\usepackage{bm}
\usepackage{tabularx}
\usepackage{algorithmic}
\newtheorem{theorem}{Theorem}
\newtheorem{definition}{Definition}

\makeatletter
\newcommand{\multiline}[1]{%
  \begin{tabularx}{\dimexpr\linewidth-\ALG@thistlm}[t]{@{}X@{}}
    #1
  \end{tabularx}
}
\makeatother

\begin{document}
%
\title{Matching Based Two-Timescale Resource Allocation for Cooperative D2D Communication}



\author{
\IEEEauthorblockN{Yiling Yuan\IEEEauthorrefmark{1}, Tao Yang\IEEEauthorrefmark{1},  Yulin Hu\IEEEauthorrefmark{2}, Hui Feng\IEEEauthorrefmark{1} and Bo Hu\IEEEauthorrefmark{1}\IEEEauthorrefmark{3}}
\IEEEauthorblockA{
\IEEEauthorrefmark{1} Research Center of Smart Networks and Systems, Fudan University, Shanghai, China\\
\IEEEauthorrefmark{2} ISEK Research Group, RWTH Aachen University, 52062 Aachen, Germany\\
\IEEEauthorrefmark{3}Key Laboratory of EMW Information (MoE), Fudan University, Shanghai, China\\
Emails: \{yilingyuan13, taoyang\}@fudan.edu.cn, hu@umic.rwth-aachen.de, \{hfeng, bohu\}@fudan.edu.cn
}}

\maketitle

\begin{abstract}
We consider a cooperative device-to-device (D2D) communication system, where the D2D transmitters (DTs) act as relays to assist cellular users (CUs) in exchange for the opportunities to use the licensed spectrum. To reduce the overhead, we propose a novel two-timescale resource allocation scheme, in which the pairing between CUs and D2D pairs is decided at a long timescale and time allocation factor for CU and D2D pair is determined at a short timescale. Specifically, to characterize the long-term payoff of each potential CU-D2D pair, we investigate the optimal cooperation policy to decide the time allocation factor based on the instantaneous channel state \mbox{information (CSI)}. We prove that the optimal policy is a threshold policy. Since CUs and D2D pairs are self-interested, they are paired only when they agree to cooperate mutually. Therefore, to study the behaviors of CUs and D2D pairs, we formulate the pairing problem as a matching game, based on the long-term payoff of each possible pairing. Furthermore, unlike most previous matching model in D2D networks, we allow transfer between CUs and D2D pairs to improve the performance. Besides, we propose an algorithm, which converges to an $\epsilon$-stable matching.
\end{abstract}


%
\IEEEpeerreviewmaketitle

\section{Introduction}
\par Recently,  device-to-device (D2D)  communication has been envisioned as a promising technology to provide a better user experience. Specifically, proximity D2D users can communicate with each other directly without going through the base station (BS). Taking the advantages of the physical proximity of communicating devices, D2D communication can improve spectrum utilization and energy efficiency, and reduce end-to-end latency. There are mainly two ways for D2D pairs to share the cellular spectrum, namely underlay and overlay D2D \cite{Asadi2014}.  In the underlay D2D communication,  cellular user (CU) and D2D pairs share the same spectrum, which incurs interference to cellular links. In contrast, overlay D2D communication allows D2D pairs to occupy dedicated spectrum, which could have been assigned to CUs. Nevertheless, in both underlay or overlay D2D, the  quality of service (QoS) of CUs will be degraded.

\par Meanwhile, CUs that are far away from BS, often suffer from poor channel quality, so that their QoS requirements are hard to meet. In this context, cooperative relay technology is thought of as a key technology to tackle this problem.  Compared to fixed relay stations that incur high expenditure, mobile user relaying is an efficient and flexible solution with low cost. 

\par Combining D2D communication and cooperative relay technology, Chen \emph{et al.} \cite{Wei2016TVT} propose a D2D-based cooperative network, where mobile devices serve as relays for CUs. However, the work does not consider any  incentive mechanism for mobile devices. In fact, mobile devices, that owned by selfish users, may be unwilling to act as relays for other devices without reward. Inspired by the idea of spectrum \mbox{leasing \cite{Pantisano2012JSAC}},  authors \mbox{in\cite{Wu2017TWC,Cao2015MWC,Chen2015WCSP,Yuan2016PIMRC}} investigate a cooperative D2D system, where the D2D transmitters (DTs) act as relays to assist CUs in exchange for the opportunities to use the licensed spectrum.  Thus, the QoS of CUs can be guaranteed and D2D pairs can obtain the transmission opportunities on licensed spectrum. As a result, a win-win situation can be achieved, which motivates CUs and D2D pairs to share the spectrum. However, above works determine the pairing between multiple CUs and multiple D2D pairs at a short timescale (e.g. at LTE scheduling time interval of 1ms), which may incur heavy signaling overhead and thus is not practical in large-scale networks.

\par In this paper, we investigate a cooperative D2D communication system, where CUs and D2D pairs cooperate with each other via spectrum leasing. Unlike previous works, in order to reduce the overhead, we propose a two-timescale resource allocation scheme. In particular, the pairing between multiple CUs and multiple D2D pairs is determined at a long timescale. On the other hand, at a short timescale, a cooperation policy allocates the transmission time for D2D link and cellular link based on the instantaneous channel state information (CSI). Under the proposed scheme, only statistic CSI is required for the pairing problem at the long timescale, while at the short timescale, the BS acquires only the instantaneous CSI between every matched CU-D2D pair to decide the transmission time. As a result, the signal overhead is significantly reduced in comparison to \cite{Wu2017TWC,Cao2015MWC,Chen2015WCSP,Yuan2016PIMRC}.

\par Moreover, we develop a matching game based framework to solve the two-timescale resource allocation problem. Specifically, we investigate the optimal cooperation policy for each D2D pair and its potential CU partner to characterize the long-term payoff of this potential pairing. In general,  CUs and D2D pairs may be of self-interest\cite{Song2014}, and thus they can only be paired when they agree to cooperate with each other.  The matching game provides an appropriate framework for such pairing problem with two-sided preferences \cite{Bayat2016MSP}. This motivates us to formulate the pairing problem at the long timescale as a one-to-one matching game, based on the long-term payoff of each potential CU-D2D pair. Furthermore, unlike previous one-to-one matching models in D2D networks \cite{Yuan2016PIMRC,Gu2015JSAC,Yuan2019TVT}, we propose to allow the transfer between CUs and D2D pairs as a performance enhancement. Then, an algorithm is proposed, which converges to an \mbox{$\epsilon$-stable} matching. 

\par The rest of this paper is organized as follows. In Section II, we introduce the system model. We study the optimal cooperation policy in Section III and investigate the pairing problem in Section IV.  Section V gives numerical results. Finally, Section VI concludes this paper.

\par \emph{Notations:} In this paper, $\mathbf{E}\{x\}$ represents the expectation \mbox{of $x$}, and $\mathbf{I}(\cdot)$ denotes the indication function.

\section{System Model}
\par We consider a single cell with a BS denoted by $b$. Because mobile devices are more likely to need help due to their limited power  budgets, we focus on uplink resource sharing. There are $M$ CUs on the cell edge suffering from poor channel conditions which could not support their QoS. At the same time, $N$ transmitter-receiver pairs are working in D2D communication mode. There is no dedicated resource allocated for D2D pairs. As a result, D2D pairs serve as relays for CUs in exchange for access to the cellular channels. In the following, we use $\mathcal{M}=\{1,2,\cdots,M\}$ and $\mathcal{N}=\{1,2,\cdots,N\}$ to denote the sets of CUs and D2D pairs, respectively. 

\par The time domain is divided into frames of fixed length. Each frame consists of $T_s$ subframe. The channel gain remains constant in each subframe and changes over different subframes. Besides, we assume that the channel gains across different subframes of the same frame are i.i.d. and follow a known distribution. At the channel occupied by CU $m$, the instantaneous channel gains of the cellular link from CU $m$ to BS, the link from CU $m$ to DT $n$, the link from DT $n$ to BS and the D2D link from DT $n$ to D2D receiver (DR) $n$ are represented as $h^m_{mb}, h^m_{mn}, h^m_{nb}, h^m_{nn}$, respectively.

\begin{figure}[!t]
\centering
\includegraphics[width=2.6in]{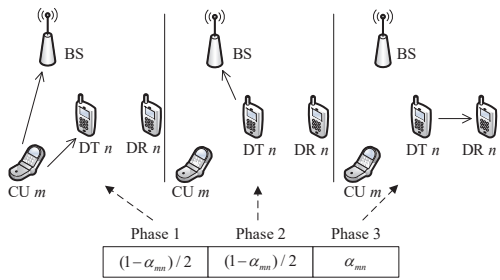}
\captionsetup{font={small}}
\caption{Subframe structure for cooperation.}
\label{systemModel}
\end{figure}

\par We assume that each CU is assisted by at most one D2D pair, since it has been shown that a single relay can achieve the full diversity gain \cite{Kadloor2010TWC}. As depicted in Fig.\ref{systemModel}, the normalized subframe is divided into three phases when D2D pair $n$ cooperates with  CU $m$.\footnote{For simplicity, we assume the transmission direction of  D2D pair is fixed during the entire frame and DT acts as a relay for CUs. In fact, our proposed scheme can be applied to a more general scenario, where  the transmission direction may change and both D2D devices can be selected as a relay.} The first two phases both last $\frac{1-\alpha_{mn}}{2}$ and are used for the relay transmission for CU $m$. Specifically, CU $m$ broadcasts its data with power $P_c$ to the BS and DT $n$ at first. Then, DT $n$ forwards the received data to the BS with power $P_d$. Besides, the first two phases can also be used for the cellular link of CU $m$ when the cellular link has better performance. The last phase lasts $\alpha_{mn}$ and is used for D2D link, where DT $n$ communicates with DR $n$ with power $P_d$. Throughout the paper, we refer to $\alpha_{mn}\in\mathcal{A}\triangleq[0,1]$ as time allocation factor for D2D link.

\par The rate of CU $m$ in the cellular link is 
\begin{equation}
r^C_{m}=\ln\left(1+\frac{P_ch^m_{mb}}{N_0}\right),
\end{equation}
where $N_0$ denotes the noise power. In this paper, we take the decode-and-forward with repetition coding as the relay scheme. Thus, when CU $m$ is aided by D2D pair $n$, the rate of CU $m$ during the first two phases is given by
\begin{equation}
r^R_{mn} = \frac{1}{2}\min\left\{\ln\left(\!1\!+\!\frac{P_ch^m_{mn}}{N_0}\!\right),\ln\left(\!1\!+\!\frac{P_ch^m_{mb}}{N_0}\!+\!\frac{P_dh^m_{nb}}{N_0}\!\right)\right\}.
\end{equation}
Since the first two phases can also be used for the cellular link of CU $m$, the achieved rate of CU $m$ during the entire subframe can be represented as
\begin{equation}
R^C_{mn}(\alpha_{mn})=(1-\alpha_{mn})\max\left\{r^C_m,r^R_{mn}\right\}.
\end{equation}
For convenience, we define $r^C_{mn}\triangleq\max\left\{r^C_m,r^R_{mn}\right\}$. 

\par At the same time, the rate of D2D pair $n$ during the entire subframe can be given as
\begin{equation}
R^D_{mn}(\alpha_{mn})=\alpha_{mn}\ln\left(1+\frac{P_dh^m_{nn}}{N_0}\right)\triangleq\alpha_{mn}r^D_{mn}.
\end{equation}

\par  Thus, we have two variables to determine: pairing between multiple CUs and multiple D2D pairs, and time allocation factor for each CU-D2D pair. To this end, we propose a matching game based framework to determine these two variables at two different timescales. Specifically, based on the instantaneous CSI, the cooperation policy decides the time allocation factor for each CU-D2D pair at each subframe (i.e. at the short timescale). We try to study the optimal cooperation policy to characterize the long-term payoff of each potential CU-D2D pair. Then, based on these long-term payoffs, we use the matching game with transfer to decide the pairing for each frame (i.e. at the long timescale).  In other words, the optimal cooperation policy is the bridge between two timescales. In the following two sections, we will study these two subproblems, respectively.

\section{Optimal Cooperation Policy}
\par In this section, we investigate the optimal cooperative policy for each CU-D2D pair. Without loss of generality,   we  assume that CU $m$ cooperates with D2D pair $n$. Define the state $\mathbf{r}_{mn}\triangleq(r^C_{mn},r^D_{mn})$, which is determined by the instantaneous CSI. The set of all the possible states $\mathbf{r}_{mn}$ is denoted by $\mathcal{R}_{mn}$. The cooperation policy decides the time allocation factor $\alpha_{mn}$ according to the current state $\mathbf{r}_{mn}$. Mathematically, the cooperation policy is a function $\pi:\mathcal{R}_{mn}\rightarrow\mathcal{A}$. Thus, given the state $\mathbf{r}_{mn}$, the rate of D2D pair $n$ and  CU $m$ can be represented as $\pi(\mathbf{r}_{mn})r^D_{mn}$ and $\left(1-\pi(\mathbf{r}_{mn})\right)r^C_{mn}$, respectively. 

\par he optimal policy aims to maximize the expected rate of the D2D pair while  guaranteeing the QoS of the CU. Therefore, the optimization problem is formulated as  
\begin{subequations}
\label{equ5}
\begin{alignat}{2}
  \max_{\pi}\quad &\mathbf{E}_{\mathbf{r}_{mn}}\left\{\pi(\mathbf{r}_{mn})r^D_{mn}\right\} \label{equ5:a}\\
  \text{s.t.}\quad & \mathbf{E}_{\mathbf{r}_{mn}}\left\{\left(1-\pi(\mathbf{r}_{mn})\right)r^C_{mn}\right\}\geq r_{th},\label{equ5:b}
\end{alignat}
\end{subequations}
where $r_{th}$ is the minimum rate requirement for the CU and the constraint (\ref{equ5:b}) is used to guarantee the QoS of  the CU. In fact, if $T_s\gg 1$, the objective function (\ref{equ5:a}) and the left-hand side of the constraint (\ref{equ5:b}) are a good approximation of the average rate of D2D pair $n$ and CU $m$ over $T_s$ subframes, respectively. In this section, all the expectations are taken over the random \mbox{variable $\mathbf{r}_{mn}$}. For brevity, we omit the subscript $\mathbf{r}_{mn}$ in the following. 

\par Next, we investigate the structure of the optimal cooperation policy in the following theorem. 

\begin{theorem}[Structure of Optimal Policy]
\label{thm1}
If the problem (\ref{equ5}) is feasible, the optimal policy $\pi^*$ is given by
\begin{equation}
\pi^*(\mathbf{r}_{mn})=\begin{cases}
0,&\quad\lambda^*r^C_{mn}>r^d_{mn},\\
\alpha^*,&\quad\lambda^*r^C_{mn}=r^d_{mn},\\
1,&\quad\lambda^*r^C_{mn}<r^d_{mn},
\end{cases}
\end{equation}
where
\begin{align}
\lambda^*&=\min\left\{\lambda|\mathbf{E}\{r^C_{mn}\mathbf{I}(\lambda r^C_{mn}\geq r^D_{mn})\}\geq r_{th}\right\},\\
\alpha^*   &=\frac{r_{th}-\mathbf{E}\{r^C_{mn}\mathbf{I}(\lambda r^C_{mn}> r^D_{mn})\}}{\mathbf{E}\{r^C_{mn}\mathbf{I}(\lambda r^C_{mn}= r^D_{mn})\}}.
\end{align}
In fact, $\lambda^*$ indicates the minimum threshold which can satisfy the constraint (\ref{equ5:b}), and $\alpha^*$ ensures the equality of  (\ref{equ5:b}).
\end{theorem}

\begin{IEEEproof}
We can construct the Lagrangian for the problem (\ref{equ5}) as follows.
\begin{align*}
\mathcal{L}(\pi,\lambda)&=\mathbf{E}\left\{\pi(\mathbf{r}_{mn})r^D_{mn}\right\}+\lambda\left(\mathbf{E}\{(1\!-\!\pi(\mathbf{r}_{mn}))r^C_{mn}\}\!-\!r_{th}\right)\\
&=\mathbf{E}\left\{\pi(\mathbf{r}_{mn})(r^D_{mn}-\lambda r^C_{mn})\right\}+\lambda\mathbf{E}\{r^C_{mn}\}-\lambda r_{th}, 
\end{align*}
where $\lambda$ is the Lagrange multiplier associated with the constraint (\ref{equ5:b}). For a fixed $\lambda$, it is easy to find out that the following \mbox{policy $\hat{\pi}_{\lambda}$}, which is given in (\ref{solutionOfLagarange}), can maximize the Lagrangian $\mathcal{L}(\pi,\lambda)$.
\begin{equation}
\label{solutionOfLagarange}
\hat{\pi}_{\lambda}(\mathbf{r}_{mn})=\begin{cases}
0,&\quad\lambda r^C_{mn}>r^D_{mn},\\
\alpha^*,&\quad\lambda r^C_{mn}=r^D_{mn},\\
1,&\quad\lambda r^C_{mn}<r^D_{mn}.
\end{cases}
\end{equation}

\par The Lagrange dual function can be given by $g(\lambda) = \max_{\pi}\mathcal{L}(\pi,\lambda)$. Thus, substituting (\ref{solutionOfLagarange}) to $g(\lambda)$, we have
\begin{equation}
\label{dual}
\begin{split}
g(\lambda) 
=& \mathcal{L}(\hat{\pi}_{\lambda},\lambda)\\
=& \mathbf{E}\left\{\mathbf{I}(\lambda r^C_{mn}\!<\!r^D_{mn})(r^D_{mn}\!-\!\lambda r^C_{mn})\right\}\!+\!\lambda\mathbf{E}\{r^C_{mn}\}\!-\!\lambda r_{th}.
\end{split}
\end{equation}

\par In the following, we show that $\lambda^*$ minimizes the Lagrange dual function.

\par Assuming $\Delta\lambda>0$. Then, using (\ref{dual}), we have
\begin{align*}
&g(\lambda^*+\Delta\lambda)-g(\lambda^*)\\
&\qquad=\Delta\lambda\mathbf{E}\{r^C_{mn}\}-\Delta\lambda r_{th}-\Delta\lambda\mathbf{E}\left\{r^C_{mn}\mathbf{I}(\lambda^*r^C_{mn}\!<\! r^D_{mn}) \right\}\\
&\qquad\quad-\mathbf{E}\Big\{\!(\lambda^*\!+\!\Delta\lambda)r^C_{mn}\mathbf{I}(\lambda^*r^C_{mn}\!\leq\! r^D_{mn\!}<\!(\Delta\lambda\!+\!\lambda^*)r^C_{mn}\!\Big\}\\
&\qquad\quad+\mathbf{E}\Big\{\!r^D_{mn}\mathbf{I}(\lambda^*r^C_{mn}\!\leq\! r^D_{mn\!}<\!(\Delta\lambda\!+\!\lambda^*)r^C_{mn}\!\Big\}\\
&\qquad\geq\Delta\lambda\mathbf{E}\{r^C_{mn}\}-\Delta\lambda\mathbf{E}\left\{r^C_{mn}\mathbf{I}(\lambda^*r^C_{mn}\!<\! r^D_{mn}) \right\}-\Delta\lambda r_{th}\\
&\qquad=\Delta\lambda\mathbf{E}\left\{r^C_{mn}\mathbf{I}(\lambda^*r^C_{mn}\!\geq\! r^D_{mn}) \right\}-\Delta\lambda r_{th}\\
&\qquad\geq 0,
\end{align*}
where the last inequality is based on the definition of $\lambda^*$.

\par On the other hand, it is easy to verify that $\hat{\pi}_{\lambda^*}(\mathbf{r}_{mn})\leq\hat{\pi}_{\lambda^*-\Delta\lambda}(\mathbf{r}_{mn})$. Consequently, using the definition of $\lambda^*$, we have the following inequalities
\begin{align*}
&g(\lambda^*)-g(\lambda^*-\Delta\lambda)\\
&\qquad\leq -\Delta\lambda\mathbf{E}\left\{\hat{\pi}_{\lambda^*}(\mathbf{r}_{mn})r^C_{mn}\right\}+\Delta\lambda\mathbf{E}\{r^C_{mn}\}-\Delta\lambda r_{th}\\
&\qquad=  -\Delta\lambda\mathbf{E}\{r^C_{mn}\mathbf{I}(\lambda^*r^C_{mn}\!<\!r^D_{mn})\}\!+\!\Delta\lambda\mathbf{E}\{r^C_{mn}\}\!-\!\Delta\lambda r_{th}\\
&\qquad=  \Delta\lambda\mathbf{E}\{r^C_{mn}\mathbf{I}(\lambda^*r^C_{mn}\geq r^D_{mn})\}-\Delta\lambda r_{th}\\
&\qquad\leq  0.
\end{align*}

\par Thus, we can conclude that $\lambda^*$ is a solution to the dual problem $\min_{\lambda\geq 0}g(\lambda)$. Therefore, we can have
\begin{equation*}
P^*\overset{(a)}\leq g(\lambda^*)=\mathcal{L}(\pi^*,\lambda^*)\overset{(b)}=\mathbf{E}\left\{\pi^*(\mathbf{r}_{mn})r^D_{mn}\right\}\overset{(c)}\leq P^*,
\end{equation*}
where $P^*$ is the optimal value of the problem (\ref{equ5}). The inequality (a) is due to the duality gap. The \mbox{equality (b)} is based on the fact that the policy $\pi^*$ can make the \mbox{constraint  (\ref{equ5:b})} hold with equality. Since $\pi^*$ is a feasible solution to the problem (\ref{equ5}), we can obtain the inequality (c). 

\par Therefore, we can conclude that $\pi^*$ is an optimal cooperation policy.
\end{IEEEproof}

\par Theorem \ref{thm1} implies that the optimal policy can be a threshold policy, which makes decisions based on the ratio of $r^D_{mn}$ to $r^C_{mn}$. Besides, this theorem also shows that this optimal policy will allocate the entire subframe for D2D transmission (i.e. $\alpha_{mn}=1$) or cellular transmission (i.e. $\alpha_{mn}=0$) in most cases. As a result, such optimal policy enables efficient implementation in practice. 

\par Furthermore, note that the term $\mathbf{E}\left\{r^C_{mn}\mathbf{I}(\lambda r^C_{mn}\geq r^D_{mn})\right\}$ increases with increasing $\lambda$. Therefore, we can use binary search to find the threshold $\lambda^*$.

\par At last, we define $v_{mn}\triangleq\mathbf{E}\left\{\pi^*(\mathbf{r}_{mn})r^D_{mn}\right\}$ if the problem (\ref{equ5}) is feasible. In the case of infeasible, we set $v_{mn} = -1$. Thus, $ v_{mn}$ can characterize the long-term payoff of D2D pair $n$ when it cooperates with CU $m$. Besides, if $v_{mn}\geq 0$, we call D2D pair $n$ being \emph{acceptable} to CU $m$. On the contrary, we call D2D pair $n$ being \emph{unacceptable} to CU $m$ when $v_{mn}< 0$.

\section{Matching Game for Pairing Problem}
\par In this section, we study the pairing problem. The assignment is represented as a binary matrix $\mathbf{X}_{M\times N}=[x_{mn}]$, where $x_{mn} = 1$ implies that CU $m$ and D2D pair $n$ are matched.  We aim to maximize the long-term sum rate of D2D pairs, which can be formulated as the following problem.
\begin{subequations}
\label{equ11}
\begin{alignat}{3}
  \max_{\mathbf{X}}\quad & \sum_{n\in\mathcal{N}}\sum_{m\in\mathcal{M}}x_{mn}v_{mn}&& \label{equ11:a}\\
  \text{s.t.}\quad & \sum_{n\in\mathcal{N}}x_{mn}\leq 1, &&\quad\forall m\in\mathcal{M},\label{equ11:b}\\
  					& \sum_{m\in\mathcal{M}}x_{mn}\leq 1,&&\quad\forall n\in\mathcal{N},\label{equ11:c}\\
  					& x_{mn}\in\{0,1\},&&\quad\forall m\in\mathcal{M},\forall n\in\mathcal{N}.\label{equ11:d}
\end{alignat}
\end{subequations}
The constraint (\ref{equ11:b}) makes sure that each CU is relayed by at most one DT. Due to the limited battery capacity, each D2D pair can relay at most one CU \cite{Wu2017TWC}, which is represented in the constraint (\ref{equ11:c}). Note that $v_{mn}=-1$ when D2D pair $n$ is unacceptable to CU $m$. Therefore, the CUs will be only matched with acceptable D2D pairs. 

\par Originally stemmed from economics \cite{Roth1990Two}, the matching theory provides a framework to tackle the  problem of pairing players in two distinct sets, based on each player's individual preference. Since the CUs and D2D pairs are self-interested, we use the matching game to characterize the cooperations between CUs and D2D pairs in the pairing problem. Moreover, Theorem 1 implies that the CU is indifferent over the acceptable D2D pairs while D2D pair may have strict preference over CUs. Therefore, we allow transfer between D2D pairs and CUs to improve the performance. Such model is called matching game with transfer\cite{Bayat2016MSP} and also referred as to assignment game\cite{Roth1990Two}. Specifically, each CU has a price charged to its matched partner. Intuitively, the price of one CU indicates the willingness of D2D pairs to cooperate with that CU. 

\begin{definition}
A \emph{one-to-one mapping} $\mu$ is a function from $\mathcal{M}\cup\mathcal{N}$ to $\mathcal{M}\cup\mathcal{N}\cup\{0\}$ such that $\mu(m)=n$ if and only if $\mu(n)=m$, and $\mu(m)\in\mathcal{M}\cup\{0\}$, $\mu(n)\in\mathcal{N}\cup\{0\}$ for $\forall m\in\mathcal{M},\forall n\in\mathcal{N}$.
\end{definition}

\par Note that $\mu(x)=0$ means that the user $x$ is unmatched in $\mu$. The above definition implies that a one-to-one mapping matches a user on one side to the one on the other side unless the user is unmatched. Thus, a mapping $\mu$ can define a feasible solution to the problem (\ref{equ11}). Next, we will introduce the price into the matching model.
\begin{definition}
A \emph{matching} is defined as $\Phi=(\mu,\mathbf{p})$, where $\mu$ is a one-to-one mapping, $\mathbf{p}=(p_1,p_2,\cdots,p_M)$ is the price vector of CUs and $p_m\geq 0,\forall m \in \mathcal{M}$. Moreover, if $\mu(m)=0$, then $p_m=0$.
\end{definition}

\par We denote the utilities of CU $m$ and D2D pair $n$ as $\theta_m$ and $\delta_n$, respectively. Thus, given a matching $\Phi=(\mu,\mathbf{p})$, $\theta_m$ and $\delta_n$ can be represented as
\begin{align}
\theta_m(\Phi)&=p_m,\\
\delta_n(\Phi) &= v_{\mu(n)n}-p_{\mu(n)}.
\end{align}
Here, we let $p_0=0$ and $v_{0n}=0,\forall n\in\mathcal{N}$ for convenience.

\par In the matching theory, the concept of \emph{stability} is important. On the other hand, since the price is usually quantized for exchange between CUs and D2D pairs in practical, we introduce the $\epsilon$-stable matching as follows. 
\begin{definition}
Given $\epsilon\geq0$, a matching $\Phi$ is \emph{$\epsilon$-stable}, if and only if the following two conditions are satisfied:
\begin{enumerate}[(1)]
\item $\theta_m(\Phi)\geq 0,\delta_n(\Phi)\geq 0$, for $\forall m\in\mathcal{M},\forall n\in\mathcal{N}$;
\item $\theta_m(\Phi)+\delta_n(\Phi)\geq v_{mn}-\epsilon$, for $\forall m\in\mathcal{M},\forall n\in\mathcal{N}$.
\end{enumerate}
\end{definition}
The condition (1) is called \emph{individual rationality condition} and reflects that a user may remain unmatched if the cooperation is not beneficial. Condition (2) implies that there is no CU-D2D pair $(m,n)$ such that they can form a new matching, where both of them can increase their utilities and one of them can improve its utility by at least $\epsilon$.

\begin{algorithm}[!t]
\caption{Algorithm to Find $\epsilon$-stable Matching}
 {\fontsize{8pt}{0.85\baselineskip}\selectfont
 \begin{algorithmic}[1]
 \renewcommand{\algorithmicrequire}{\textbf{Initialization:}}
 \STATE Set $t=1, p_m=\beta_m^t=0,\mu^0(m)=0,\forall m\in\mathcal{M}$;
  \renewcommand{\algorithmicrequire}{\textbf{ D2D Pairs' Proposals:}}
 \REQUIRE
 \STATE  Broadcast the price requirement vector $\boldsymbol{\beta}^t=({\beta}^t_1,\beta^t_2,\cdots,\beta^t_M)$;
 \FOR{each unmatched D2D pair $n\in\mathcal{N}$}
 \STATE  Determine its demand $m=D_n(\bm{\beta}^t)$;
  \STATE  { If $m\neq 0$, D2D pair $n$ proposes to CU $m$ \mbox{($g^t_{mn} = 1$)}.  Otherwise, D2D pair $n$ does not proposes ($g^t_{mn} = 0,\forall m\in\mathcal{M}$) and $\mu^t(n)=0$;}
\ENDFOR
 \renewcommand{\algorithmicrequire}{\textbf{ CUs' Decision Making:}}
\REQUIRE
 \FOR{Each CU $m\in\mathcal{M}$}
 \IF{$\sum_{n\in\mathcal{N}}g^t_{mn}\!=\!0$, $\sum_{n\in\mathcal{N}}g^{t-1}_{mn}>0$ and $\mu(m)=0$}
  \STATE  Set $\mu^t(m)=n^*$, where $n^*=random(\{n|g^{t-1}_{mn}\!=\!1\})$;
 \STATE Set $p_m=\beta^{t-1}_m$ and $\beta^{t+1}_m = \beta^{t}_m$;
  \STATE  Set $g^t_{m^*n^*}=0$, where $m^* = D_{n^*}(\bm{\beta}^t)$;
 \ENDIF
 \ENDFOR
 \FOR {Each CU $m\in\mathcal{M}$}
 \IF {$\sum
 \limits_{n\in\mathcal{N}}g^t_{mn}\!=\!1$, and $\mu^{t-1}_m\!=\!0$ or $p_m\!<\!\beta^t_m$ are satisfied }
  \STATE Set $\mu^{t}(m)=n^*$, where $g^t_{mn^*}=1$;
  \STATE Set $p_m=\beta^t_m$ and $\beta^{t+1}_m=\beta^t_m$;
\ELSIF{$\sum_{n\in\mathcal{N}}g^t_{mn}\geq1$}
   \STATE  Set $\mu^t(m)=0$;
   \STATE  If $n\!\neq\!0$ and $p_m\!=\!\beta^t_m$ where $n\!=\!\mu^{t-1}(m)$, set $g^t_{mn}\!=\!1$;
  \STATE  Set $\beta^{t+1}_m=\beta^t_m+\epsilon$;
 \ELSE
  \STATE Set $\beta^{t+1}_m=\beta^t_m$;
 \ENDIF
 \ENDFOR
  \STATE  $t\leftarrow t+1$;
 \STATE  Go to step 3 until there is no proposal in current loop.
 \end{algorithmic}}
 \end{algorithm}
 
 \par  Algorithm 1 is proposed to find an $\epsilon$-stable matching. In the following, we give a brief description of the algorithm during $t$-th iteration.
 \par At first, the price requirement vector $\bm{\beta}^t$ will be broadcast, and $\beta^t_m$ represents the minimum price has to pay if D2D pair wants to propose to CU $m$ at the current iteration. Then, each unmatched D2D pair $n$ selects its favorite CU according to $D_n(\bm{\beta}^t)$, where the demand function is represented as 
 \begin{equation}
 D_n(\bm{\beta}^t)=\begin{cases}
 arg\max\limits_ {m\in\mathcal{M}}(v_{mn}-\beta_m^t),&\max\limits _{m\in\mathcal{M}}(v_{mn}-\beta_m^t)\geq0 ,\\
 0,&\text{otherwise.}
 \end{cases}
 \end{equation}
 
 \par In the CUs' decision making stage, the CUs decide if they want to match with the D2D pairs. There are four cases for each CU $m$. The first case (step 8-12) is that CU  $m$ is unmatched and receives no proposals after increasing its price requirement, but in the previous iteration, has received proposals from multiple D2D pairs. Then, CU $m$ will select randomly one of those D2D pairs to be matched with and set the price as $p_m=\beta^{t-1}_m$. The second case (step 15-17) is that CU $m$ receives one proposal, and meanwhile, it is unmatched or matched with the price $p_m<\beta^t_m$. In other words, only one D2D pair wants to be matched with CU $m$ with the price $\beta^t_m$.  As a result, CU $m$ will be matched with that D2D pair and set the price as $p_m=\beta^t_m$. The third  case (step 18-21) is that there are multiple D2D pairs (including the current partner of CU $m$) wanting to be matched with CU $m$ with the price $\beta^t_m$. Then,  CU $m$ will increase its price requirement by $\epsilon$ and become unmatched, where $\epsilon> 0$ is the price-step.  In the last \mbox{case (step 23)}, CU $m$ will remain the price requirement and do nothing else. The convergence of Algorithm 1 is given in the following theorem.
\begin{theorem}
The Algorithm 1 converges to an $\epsilon$-stable matching.
\end{theorem}

\begin{IEEEproof}
At first, we show that the algorithm converges to a matching. Note that $\beta^t_m$ is non-decreasing. Moreover, it can be found that $\beta^t_m\leq\epsilon+\max_{n\in\mathcal{N}}v_{mn}$. Therefore, the algorithm will converge in finite steps. Use $\Phi=(\mu,\mathbf{p})$ to denote the final result. It is easy to verify that once a CU has received a proposal,  the CU will have a partner in $\mu$. Thus, the prices of the CUs unmatched in $\mu$ are zero. Therefore, $(\mu,\mathbf{p})$ is a matching.

\par In the following, we prove that $(\mu,\mathbf{p})$ is $\epsilon$-stable by contradiction. 

\par Suppose there exists a CU-D2D pair $(m,n)$ such that $\theta_m+\delta_n< v_{mn} -\epsilon$. Assume $\mu(m)=n'$ and $\mu(n)=m'$. Thus, we can find a price $p'$ such that $p'\geq p_m+\epsilon$ and $v_{mn}-p'>v_{m'n}-p_{m'}$. So, we have
\begin{equation}
\label{equ14}
v_{mn}-p_m>v_{mn}-p'>v_{m'n}-p_{m'}.
\end{equation}
According to the algorithm, (\ref{equ14}) implies that D2D pair $n$ must have proposed to CU $m$. Therefore, there must exist an iteration, denoted by $\tau$-th iteration, where the first case happens for CU $m$. Specifically, CU $m$ receives multiple proposals with $\beta^{\tau-1}_m=p_m$ at $(\tau-1)$-th iteration, and receives no proposal with $\beta^{\tau}_m=p_m+\epsilon$  at  $\tau$-th iteration. Furthermore, no D2D pairs propose to CU $m$ afterward. As a result, we have the following inequalities
\begin{equation}
v_{mn}-p'\leq v_{mn}-\beta^\tau_m<v_{m'n}-p_{m'},
\end{equation}
which is inconsistent with (\ref{equ14}). 
\par Besides, it is easy to verify $(\mu,\mathbf{p})$ satisfies the individual rationality condition. Therefore, we can conclude that $(\mu,\mathbf{p})$ is $\epsilon$-stable.
\end{IEEEproof}

\section{Simulation Results}
In this section, the performance of the proposed scheme is investigated through simulations. The instantaneous channel gain used in the simulation is $h=\eta L^{-\gamma}$, where $\eta$ is the fast-fading gain with exponential distribution, $\gamma=4$ is the pathloss exponent and $L$ is the distance between transmitter and receiver. For simulation, we consider the scenario where the BS is deployed in the cell center while the radius of the cell is set to 500 m. The CUs are distributed uniformly at the cell edge. Meanwhile, the D2D pairs are uniformly distributed in the area with a distance of 200 m to 400 m from the BS. Other configuration parameters are given in Table.I.

\begin{table}[!t]
\footnotesize
\caption{Configuration Parameters}
\renewcommand{\arraystretch}{1}
\centering
\begin{tabular}{|c|c|}
\hline
\bf{Parameters} & \bf{Value}\\
\hline
Power noise ($N_0$) & -100 dBm\\
\hline
Transmit power of CU ($P_c$) & 20 mW\\
\hline
Transmit power of DT ($P_d$) & 20 mW\\
\hline
Distance of D2D link & Uniformly distributed in $[10, 30]$ m\\
\hline
Minimum rate requirement ($r_{th}$) & 1.8 bps/Hz\\
\hline
Price-step ($\epsilon$) & 1\\
\hline
\end{tabular}
\end{table}

\begin{figure}[!t]
\centering
\captionsetup{font={small}}
\includegraphics[width=2.8in]{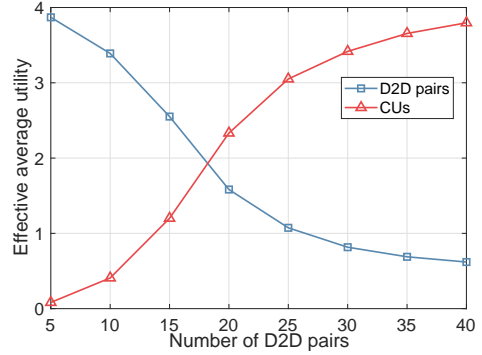}
\caption{The effective average utilities of D2D pairs and CUs versus the number of D2D pairs, where $M=15$.}
\label{AvgUtility}
\end{figure}

\par At first, we investigate the property of our scheme. Fig.\ref{AvgUtility} presents the effective average  utilities (EAU) of CUs and D2D pairs versus the number of D2D pairs. The EAU of CUs is defined as $\text{EAU} = \frac{\text{Sum of CUs' utilities}}{\text{Number of matched CUs}}$. The EAU of D2D pairs can be defined in a similar way. It can be observed that with the increasing number of D2D pairs, the EAU of CUs increases while the EAU of D2D pairs decreases. The rationale behind this is that when $N$ is small, there is a strong competition among CUs to acquire the relay service from D2D pairs. Therefore, the prices of CUs are low and each matched D2D pair has high utility. In comparison, when there is a large number of D2D pairs, the available CUs become a scarce resource. As a result, each D2D pair has to pay a higher price for the transmission opportunities on the cellular channels.

\begin{figure}[!t]
\centering
\captionsetup{font={small}}
\includegraphics[width=2.8in]{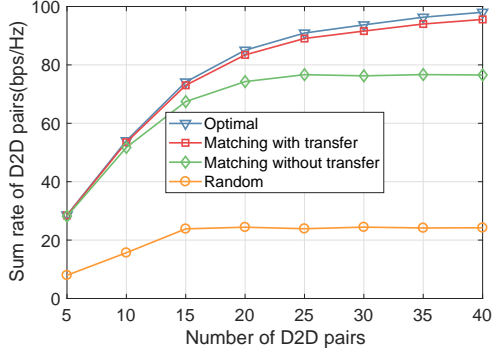}
\caption{The sum rate of D2D pairs with different schemes versus the number of D2D pairs, where $M=15$.}
\label{SumRateD2D}
\end{figure}

\begin{figure}[!t]
\centering
\captionsetup{font={small}}
\includegraphics[width=2.8in]{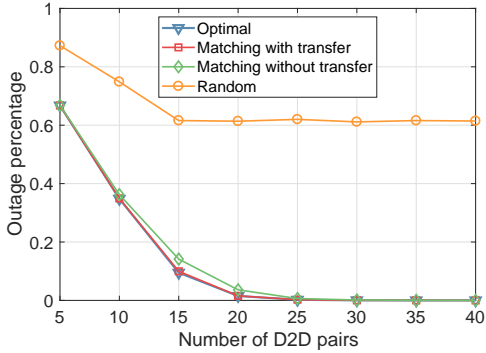}
\caption{The outage percentage of CUs with different schemes versus the number of D2D pairs, where $M=15$.}
\label{Outage}
\end{figure}

\par Next, to evaluate the performance gain of the proposed scheme, we compare it with the following schemes: i) the \emph{optimal} scheme provides the optimal solution to the problem  (\ref{equ11}); ii) the  \emph{matching without transfer} scheme adopts the matching game without transfer to solve the pairing problem (\ref{equ11}), i.e. the prices of CUs are zero; iii) the \emph{random} scheme matches the D2D pairs and CUs randomly. The comparison results are provided in Fig.\ref{SumRateD2D} and Fig.\ref{Outage}.

\par In Fig.\ref{SumRateD2D}, we compare the performance of different schemes in terms of the sum rate of D2D pairs. This figure shows that the proposed scheme achieves near-optimal performance. Besides, owning to allowing transfer between CUs and D2D pairs, the proposed scheme outperforms the matching without transfer scheme, especially in the large $N$ region. It also can be observed that the gain is small in the small $N$ region. This is due to the fact the prices of CUs are close to zero when $N$ is small (which is consistent with the results in Fig.\ref{AvgUtility}). Thus, these two schemes are almost the same in this situation. Moreover, the matching without transfer scheme only matches the CUs with their acceptable partners. Therefore, this scheme can obtain better performance than the random scheme.

\par Fig.\ref{Outage} presents the outage percentage of CUs under different schemes, and the outage refers to the case where the rate requirement of a CU is not satisfied. Compared with the random scheme, the rest three schemes achieve significantly better performance. In particular, these three schemes have a similar outage percentage. The explanation is as follows. The outage never happens if each CU is matched with acceptable D2D pairs. Since all the three schemes only match the CUs with their acceptable D2D pairs, the price has little impact on the outage percentage. Furthermore, when $N\geq20$, the outage percentage of our scheme is close to zero. It implies that our scheme  improves the performance of CUs greatly. On the contrary, the outage percentage of the random scheme is larger than 60\%, which indicates that it is essential to have an efficient pairing between CUs and D2D pairs.

\section{Conclusion}
In this paper, we have investigated a cooperative D2D communication system, where D2D pairs and CUs cooperate with each other via spectrum leasing. We have provided a low-overhead design for the system by proposing a two-timescale resource allocation scheme, in which the pairing between CUs and D2D pairs is decided at the long timescale and time allocation factor is determined at the short timescale. Specifically, to characterize the long-term payoff of each potential CU-D2D pair, we investigate the optimal cooperation policy to decide the time allocation factor. Based on these long-term payoffs, we use the matching game with transfer to solve the pairing problem. The simulation results confirm the performance gain of the proposed scheme.

\section*{ACKNOWLEDGMENT}
This work was supported in part by the National Key Research and Development Program of China No.213 and in part by the National Science Foundation of China under Grants 71731004 and Shanghai Municipal Natural Science Foundation under Grants 19ZR1404700.

\bibliographystyle{IEEEtran}
\bibliography{ref.bib}

\end{document}